# A semi-analytical method to estimate the slip length of spreading cap-shaped droplets using Cox theory

*Martin Wörner,*[*,1] *Xuan Cai*[1]*, Hocine Alla*[2] *and Pengtao Yue*[3]

[1] Karlsruhe Institute of Technology (KIT), Institute of Catalysis Research and Technology, Engesserstr. 20, 76131 Karlsruhe, Germany

[2] Université des Sciences et de la Technologie d'Oran, BP 1505 El M'Naouar Bir el Djir 31000, Oran, Algeria

[3] Virginia Tech, Department of Mathematics, 460 McBryde Hall, Blacksburg, VA 24061-0123, USA

ABSTRACT

The Cox-Voinov law on dynamic spreading relates the difference between the cubic values of the apparent contact angle $(\theta)$ and the equilibrium contact angle to the instantaneous contact line speed $(U)$. Comparing spreading results with this law requires accurate data of $\theta$ and $U$ during the entire process. We consider the case when gravitational forces are negligible and transform the general Cox law in a relationship for the temporal evolution of the spreading radius. For cap-shaped droplets, this enables a comparison of experimental or computational results with Cox theory without the need for instantaneous data of $\theta$ and $U$. The fitting of Cox theory against measured or computed base-radius-over-time curves allows estimating the effective slip length. This is useful for establishing relationships between slip length and parameters in numerical methods for moving contact lines. The procedure is illustrated by numerical simulations for partially wetting droplets employing the coupled level-set volume-of-fluid and phase field methods.

# 1  INTRODUCTION

A common approach for describing dynamic wetting and spreading processes of droplets on solid substrates is relating the macroscopic apparent contact angle ($\theta$) with the static (equilibrium) contact angle ($\theta_e$) and the contact line speed ($U$). Different forms of this relationship have been proposed by Hoffman[1], Voinov[2], Tanner[3] and Cox[4]. According to the latter reference, it is

$$G(\theta,\eta) - G(\theta_e,\eta) = Ca \ln(L/L_S) \tag{1}$$

where

$$G(\theta,\eta) = \int_0^\theta \frac{d\theta}{f_{Cox}(\theta,\eta)} \tag{2}$$

and

$$f_{Cox}(\theta,\eta) = \frac{2\sin\theta\left\{\eta^2(\theta^2 - \sin^2\theta) + 2\eta\left[\theta(\pi-\theta) + \sin^2\theta\right] + \left[(\pi-\theta)^2 - \sin^2\theta\right]\right\}}{\eta(\theta^2 - \sin^2\theta)\left[(\pi-\theta) + \sin\theta\cos\theta\right] + \left[(\pi-\theta)^2 - \sin^2\theta\right](\theta - \sin\theta\cos\theta)} \tag{3}$$

Here, $\eta = \mu_G / \mu_L$ is the gas-to-drop viscosity ratio. The capillary number $Ca := \mu_L U / \sigma$ represents a non-dimensional contact line speed normalized by the drop viscosity ($\mu_L$) and surface tension ($\sigma$). The argument of the logarithm in Eq. (1) is given by the ratio between a characteristic macroscopic length scale ($L$) and the slip length ($L_S$). The slip length is, in practice, a fitting parameter and represents the length of the region where no-slip boundary conditions do not apply[5]. The magnitude of the term $\lambda := \ln(L/L_S)$ is expected to be[6] $\sim 10$. For $\theta < 135°$ and very small viscosity ratio ($\eta \to 0$) it is

$$G(\theta,0) - G(\theta_e,0) \approx (\theta^3 - \theta_e^3)/9 \tag{4}$$

Then, Eq. (1) simplifies to the form

$$\theta^3 = \theta_e^3 + 9Ca\ln(L/L_s) = \theta_e^3 + 9\lambda Ca \qquad (5)$$

which is commonly known as the Cox-Voinov law. Experiments[7, 8] confirm that the cubic relation $\theta^3 - \theta_e^3 \sim Ca$ holds for contact angles as large as $70°-100°$ provided $Ca$ and the Reynolds number $Re := \rho_L LU/\mu_L$ are sufficiently smaller than unity.

The principle result in experimental and computational studies on dynamic spreading processes is the time evolution of the radius of the circular contact area $a(t)$ (spreading radius). Often a comparison of the experimental/numerical results with theoretical or empirical relations is of interest, e.g. for model testing or code validation. The comparison with the Cox-Voinov law, Eq. (5), is often performed by displaying $\theta^3 - \theta_e^3$ over $Ca$ (see e.g. Pahlavan et al.[9] for a recent numerical example). For experiments, this procedure is associated with two disadvantages. First, it requires the measurement or evaluation of $\theta(t)$ during the entire spreading process; this is elaborate and measurements of the apparent contact angle may be ambiguous. Second, it is potentially inaccurate because the contact line speed $U(t) = da(t)/dt$ (spreading speed) is obtained by differentiation, and thus may lead to scattered data.

In this paper, an alternative routing is presented for comparing results of dynamic droplet spreading with the Cox or Cox-Voinov law. We consider the case when gravitational effects are negligible (i.e. the Eötvös number is sufficiently small) so that the spreading is driven by capillary forces alone and the interface forms (at each instant in time) a spherical cap. Using this assumption, we transform Eq. (1) and (5) into a relationship $a(t)$ which allows easy and straightforward comparisons with experiments and computations. Furthermore, the method allows extracting effective slip length from measured or computed base-radius-over-time curves.

To illustrate the procedure, we perform numerical simulations of the spreading of a small droplet on a partially wetting flat substrate using a coupled level-set volume-of-fluid method and a phase-field method and compare the respective spreading dynamics with Cox theory.

## 2 THEORY

In this section, we transform the general Cox law into a time dependent relation for the spreading radius. For this purpose, we consider a drop with constant volume ($V$) that spreads on a flat chemically homogenous surface. We assume that during the entire spreading process the drop takes on the shape of a spherical cap. This assumption is reasonable supposed the Eötvös number is sufficiently low. By geometrical constraints, the spreading radius $a(t)$ and the apparent contact angle $\theta(t)$ are related as[10, 11]

$$a(t) = \left(\frac{3V}{\pi}\right)^{1/3} f_{\text{geo}}(\theta) \tag{6}$$

where

$$f_{\text{geo}}(\theta) := \frac{\sin\theta}{\left(2 - 3\cos\theta + \cos^3\theta\right)^{1/3}} \tag{7}$$

The time derivative of Eq. (6) is

$$\frac{da(t)}{dt} = \left(\frac{3V}{\pi}\right)^{1/3} f'_{\text{geo}}(\theta) \frac{d\theta}{dt} \tag{8}$$

where

$$f'_{\text{geo}}(\theta) = \frac{df_{\text{geo}}}{d\theta} = -\frac{1}{(2+\cos\theta)\left(2-3\cos\theta+\cos^3\theta\right)^{1/3}} \tag{9}$$

With the identity $2 - 3\cos\theta + \cos^3\theta = (1-\cos\theta)^2(2+\cos\theta)$, Eq. (8) becomes equivalent to Eq. (4.2) in Voinov[2].

By introducing the definition $Ca = \mu_L (da/dt)/\sigma$ into Eq. (1), the Cox law can be rewritten in the form

$$\frac{da(t)}{dt} = \frac{\sigma}{\mu_L \lambda}\left[G(\theta,\eta) - G(\theta_e,\eta)\right] \quad (10)$$

Combining Eq. (8) and Eq. (10) yields

$$\frac{\sigma}{\mu_L \lambda}\left(\frac{\pi}{3V}\right)^{1/3} dt = \frac{f'_{\text{geo}}(\theta)}{G(\theta,\eta) - G(\theta_e,\eta)} d\theta \quad (11)$$

An appropriate length scale for transferring Eq. (11) into a non-dimensional form is the volume-equivalent drop radius $R_V := (3V/4\pi)^{1/3}$. Using the capillary-viscous time scale $t_{\text{ref}} := \mu_L R_V / \sigma$, the non-dimensional time $\tau := t/t_{\text{ref}} = \sigma t / \mu_L R_V$ can be defined. With these definitions Eq. (11) becomes

$$d\tau = \frac{\sqrt[3]{4}\lambda f'_{\text{geo}}(\theta)}{G(\theta,\eta) - G(\theta_e,\eta)} d\theta \quad (12)$$

Eq. (12) constitutes a differential relation between the instantaneous macroscopic contact angle $\theta$ and the non-dimensional time $\tau$.

Integrating Eq. (12) requires the specification of initial conditions. Let $\theta_0 \neq \theta_e$ be the initial contact angle at time $t = \tau = 0$. Then, the initial spherical cap radius is

$$R_{S,0} = \left(\frac{3}{\pi}\frac{V}{2 - 3\cos\theta_0 + \cos^3\theta_0}\right)^{1/3} \quad (13)$$

and the initial base radius $a_0 = R_{S,0} \sin\theta_0$. For $\theta_0 > \theta_e$, the contact line advances as the droplet spreads out and wets the substrate, whereas it recedes for $\theta_0 < \theta_e$ where dewetting occurs.

Integrating Eq. (12) and taking into account Eq. (2) and Eq. (9) yields

$$\tau = -\int_{\theta_0}^{\theta} \left[ \int_0^x \frac{dy}{f_{Cox}(y,\eta)} - \int_0^{\theta_e} \frac{dy}{f_{Cox}(y,\eta)} \right]^{-1} \frac{\sqrt[3]{4}\lambda}{(2+\cos x)(2-3\cos x+\cos^3 x)^{1/3}} dx \quad (14)$$

In the limit $\eta \to 0$ Eq. (14) simplifies – by virtue of Eq. (4) – to the form

$$\tau = -\int_{\theta_0}^{\theta} \frac{9}{x^3 - \theta_e^3} \frac{\sqrt[3]{4}\lambda}{(2+\cos x)(2-3\cos x+\cos^3 x)^{1/3}} dx \quad (15)$$

The integrals on the right-hand-side of Eq. (14) and (15) cannot be solved analytically. However, for a given value of $\theta$ one can solve either integral numerically. Doing this for a set of distinct values $\theta_i$ in the range $\theta_e < \theta_i \leq \theta_0$ (spreading case) yields the corresponding set of discrete values of non-dimensional time $\tau_i$. From the discrete values $\theta_i$, one obtains from Eq. (6) the corresponding discrete values of the spreading radius $a_i$, and from the discrete values of $\tau_i$ the discrete values of $t_i$. Thus, a relation between $a_i$ and $t_i$ is established which can be used to compare the spreading dynamics of Cox theory to experimental or computational results.

Due to the lack of analytical solutions for the integrals in Eq. (14) and Eq. (15), each combination of $\theta_e$, $\theta_0$ and $\eta$ requires a separate numerical integration. This numerical solution applies by virtue of the non-dimensional time to different liquid properties and initial drop volumes. To perform the numerical integration, a MATLAB script is used (see supplemental material). Here, we choose the step size $\delta\theta = 0.001$ for the contact angle and define $\theta_i = \theta_0 - i \cdot \delta\theta$, where $i = 1, 2, 3, \ldots$. The procedure for dewetting ($\theta_e > \theta_0$) is similar and the MATLAB script can easily be adapted to handle this case as well.

# 3  NUMERICAL SIMULATION

In this section, numerical simulations of the spreading of a liquid droplet on a perfectly smooth, chemically homogenous, solid surface are presented. The simulations are performed by two different numerical methods, namely a coupled level-set volume-of-fluid (CLSVOF) method and a phase-field (PF) method. For the CLSVOF method, the implementation in the commercial code ANSYS FLUENT 15.0 is used. The PF method is implemented in the finite volume open source code OpenFOAM® (denoted as phaseFieldFoam[12, 13, 14], or in short PFF) and in an in-house finite element code (denoted as AMPHI[15, 16]).

## 3.1  Coupled Level-Set Volume-of-fluid method

The coupled level set (LS) and volume-of-fluid (VOF) method (CLSVOF) combines the advantages of both interface-capturing approaches. The method was introduced by Bourlioux[17] and became popular with the works of Sussman and Puckett[18] and Son and Hur[19]. In the LS method, the interface $\Gamma = \{\mathbf{x} \,|\, \varphi(\mathbf{x},t) = 0\}$ is implicitly defined as the zero level set of a scalar function. The LS function $\varphi$ is the signed distance from the interface, which is positive in the liquid (L) phase and negative in the gas (G) phase. Thus, $\varphi$ is smooth and its gradient can accurately be evaluated for computing the interface normal vector and curvature, cf. Eq. (26). This is a distinct advantage as compared to the VOF method where the liquid volumetric fraction $F$ (with $F = 0$ and $F = 1$ in pure gas and liquid cells, respectively, and $0 < F < 1$ otherwise) exhibits large gradients close to the interface.

The VOF and LS functions are each advanced in time by an advection equation

$$\frac{\partial F}{\partial t} + \nabla \cdot (\mathbf{u} F) = 0, \qquad \frac{\partial \varphi}{\partial t} + \nabla \cdot (\mathbf{u} \varphi) = 0 \tag{16}$$

where the velocity field **u** is determined from the Navier-Stokes equation (see below). The VOF equation is solved in a geometric manner (employing geometric piecewise linear interface construction according to Youngs[20]) which ensures good mass conservation. In contrast to $F$, the LS function $\varphi$ is not a conserved quantity. A re-initialization of the distance function is therefore performed at each time step to reduce mass conservation errors.

The modelling of wall adhesion follows Brackbill et al.[21] where $\theta_e$ is not specified as a boundary condition at the wall, but is used to adjust the interface normal vector $\hat{\mathbf{n}}_\Gamma$ in cells near the wall

$$\hat{\mathbf{n}}_\Gamma = \hat{\mathbf{n}}_w \cos\theta_e + \hat{\mathbf{t}}_w \sin\theta_e \tag{17}$$

This so-called dynamic boundary condition results in the adjustment of the curvature of the interface near the wall. Here, $\hat{\mathbf{n}}_w$ is the unit vector normal to the wall (pointing into the fluids) and $\hat{\mathbf{t}}_w$ is the unit vector tangential to the wall.

### 3.2 Phase-field method

In the phase-field method[22, 23, 24, 25, 26], the order parameter $C$ serves to describe the distribution of the gas and liquid phases. Here, $C$ takes distinct values $C_L = 1$ and $C_G = -1$ in the bulk phases and varies rapidly but smoothly in a thin transition layer (the diffuse interface). The location of the gas-liquid interface is represented by $C = 0$. To determine the phase evolution, the convective Cahn-Hilliard equation

$$\frac{\partial C}{\partial t} + \nabla \cdot (\mathbf{u}C) = \kappa_C \nabla^2 \phi \tag{18}$$

is solved, where

$$\phi(C) = \frac{3}{2\sqrt{2}} \frac{\sigma}{\varepsilon} \left[ C(C^2 - 1) - \varepsilon^2 \nabla^2 C \right] \tag{19}$$

is the chemical potential. In the latter equations, $\kappa_C$ is the mobility parameter and $\varepsilon$ is a positive constant that is related to the interfacial thickness. In equilibrium, the variation $-0.9 \leq C \leq 0.9$ occurs over a distance of about $4.1\varepsilon$. The boundary condition for the order parameter is[23]

$$\hat{\mathbf{n}}_w \cdot \nabla C = \frac{\sqrt{2}}{2} \frac{\cos\theta_e}{\varepsilon}(1-C^2) \tag{20}$$

Here, the interfacial width parameter $\varepsilon$ and the mobility $\kappa_C$ are considered as numerical rather than as physical parameters. Their values are quantified via two appropriate dimensionless parameters. The first one is the Peclet number $Pe_C := \sqrt{8/9} R_V U_{ref} \varepsilon /(\kappa\sigma)$ for the order parameter. Here, we use the capillary-viscous velocity as reference velocity ($U_{ref} = \sigma/\mu_L$) so that $Pe_C = \sqrt{8/9} R_V \varepsilon /(\kappa\mu_L)$. The Cahn number $Cn := \varepsilon / R_V$ relates the interfacial width parameter to the volume-equivalent drop radius.

### 3.3 Navier-Stokes equation

We consider two incompressible, immiscible, isothermal Newtonian fluids with constant physical properties. The flow of both phases is described by the single-field Navier-Stokes equations

$$\nabla \cdot \mathbf{u} = 0 \tag{21}$$

$$\rho\left(\frac{\partial \mathbf{u}}{\partial t} + \mathbf{u} \cdot \nabla \mathbf{u}\right) = -\nabla p + \nabla \cdot \mu\left[\nabla \mathbf{u} + (\nabla \mathbf{u})^T\right] + \rho \mathbf{g} + \mathbf{f}_{st} \tag{22}$$

The density field and viscosity field depend on the phase distribution; they are computed as

$$\rho(\mathbf{x},t) = \rho_G(1-H) + \rho_L H, \qquad \mu(\mathbf{x},t) = \mu_G(1-H) + \mu_L H \tag{23}$$

where $H(\mathbf{x},t)$ is a Heaviside function. In the CLSVOF method, the Heaviside function

$$H_b(\varphi) = \begin{cases} 0 & \text{if } \varphi < -b \\ \frac{1}{2}\left[1 + \frac{\varphi}{b} + \frac{1}{\pi}\sin\left(\frac{\pi\varphi}{b}\right)\right] & \text{if } |\varphi| \leq b \\ 1 & \text{if } \varphi > b \end{cases} \quad (24)$$

is smoothed over a distance $2b$ where $b = 1.5h$ is related to the grid spacing $h$. In the PF method, the Heaviside function is $H_C = (1+C)/2$.

The last term in Eq. (22) is the surface tension term. In the CLSVOF method, it is modelled as

$$\mathbf{f}_{st} = -\sigma\kappa\delta_a(\varphi)\nabla\varphi \quad (25)$$

where the interface curvature is computed as

$$\kappa = \nabla \cdot \left(\frac{\nabla\varphi}{|\nabla\varphi|}\right) \quad (26)$$

The smoothed delta function $\delta_b$ is obtained from the LS function as

$$\delta_b(\varphi) = \frac{dH_b(\varphi)}{d\varphi} = \begin{cases} \frac{1+\cos(\pi\varphi/a)}{2b} & \text{if } |\varphi| < b \\ 0 & \text{if } |\varphi| \geq b \end{cases} \quad (27)$$

In the PF method, different formulations of the surface tension force are used in the literature.[26] In phaseFieldFoam the surface tension term is computed as

$$\mathbf{f}_{st} = -C\nabla\phi = -\frac{3}{2\sqrt{2}}\frac{\sigma}{\varepsilon}C\nabla\left[C(C^2-1) - \varepsilon^2\nabla^2 C\right] \quad (28)$$

whereas in AMPHI

$$\mathbf{f}_{st} = \phi\nabla C \quad (29)$$

is used. The two formulations end up with different pressure levels but this does not matter for incompressible flows.

## 3.4 Computational set-up and fluid properties

Figure 1 shows the polar coordinate systems and a sketch of the initial and equilibrium droplet shapes. The initial shape is a hemisphere so that the initial contact angle is $\theta_0 = 90°$. The initial spherical cap radius is $R_{S,0} = 0.5\,\text{mm}$. This corresponds to a drop volume $V = 0.2618\,\text{mm}^3$ and volume-equivalent drop radius $R_V = 0.397\,\text{mm}$. The liquid phase is coconut oil with a density $\rho_L = 910\,\text{kg/m}^3$ and dynamic viscosity $\mu_L = 0.03\,\text{Pa}\,\text{s}$. The gas density is $\rho_G = 1\,\text{kg/m}^3$ and the coefficient of surface tension is $\sigma = 0.0294\,\text{N/m}$. The value of the Eötvös number $Eo := (\rho_L - \rho_G) g R_V^2 / \sigma = 0.0478$ is much smaller than unity. Thus, gravitational effects are small and the spreading is driven by capillarity. In the simulations, gravity is therefore neglected ($g = 0$) if not mentioned otherwise. The value of the capillary-viscous time scale is $t_{\text{ref}} = 0.405\,\text{ms}$.

The 2D computational domain is a square with size $0 \le r \le H$ and $0 \le z \le H$. The boundary conditions are as follows. At the axis ($r = 0$) axi-symmetry is specified. At the flat and smooth solid surface ($z = 0$) the no-slip condition holds. At the top and right boundaries of the computational domain ($r = z = H$) constant pressure inlet boundary conditions are applied. In the simulations with the finite volume codes Fluent and phaseFieldFoam, a uniform Cartesian grid with mesh size $\Delta z = \Delta r = h$ is used. The numerical simulations in this paper are all performed with a fixed equilibrium contact angle that is independent from the contact line speed. However, the apparent contact angle determined at a certain distance away from the wall may be different from $\theta_e$ (cf. **Figure 4** below).

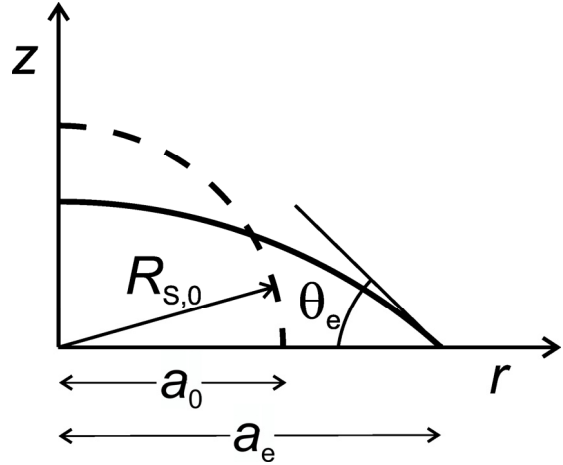

**Figure 1.** Schematic illustration of the polar coordinate system with initial drop shape (dashed line, contact angle $\theta_0 = 90°$) and final drop shape (solid line, equilibrium contact angle $\theta_e < \theta_0$).

## 4   RESULTS AND DISCUSSION

In this section, we compare numerical results for the time evolution of the spreading radius with the respective curve derived from Cox theory using the MATLAB script. Two different cases are considered. In the first case, results of the two finite volume based codes Fluent and phaseFieldFoam are compared with the Cox theory. In the second case, the effect of inertia is studied with the finite element code AMPHI.

### 4.1   Influence of viscosity ratio

We first discuss the influence of the viscosity ratio ($\eta$) on the motion of the contact line for the general Cox law. Since Eq. (14) is linear with respect to $\lambda = \ln(L/L_s)$, it is sufficient to consider one value only (here $\lambda = 10$). In **Figure 2** we show results for four different values of $\eta$, namely 0, 0.01, 0.1 and 1. It is obvious that the spreading curves for $\eta = 0$ and $\eta = 0.01$ are almost identical. This indicates that the influence of the gas viscosity on the spreading dynamics is negligible for $\eta \leq 0.01$. Note that the curve for $\eta = 1$ in **Figure 2** (which results from general Cox theory) is not consistent with the Cox-Voinov law, Eq. (5), since Eq. (4) is only valid in the limit of very small viscosity ratio.

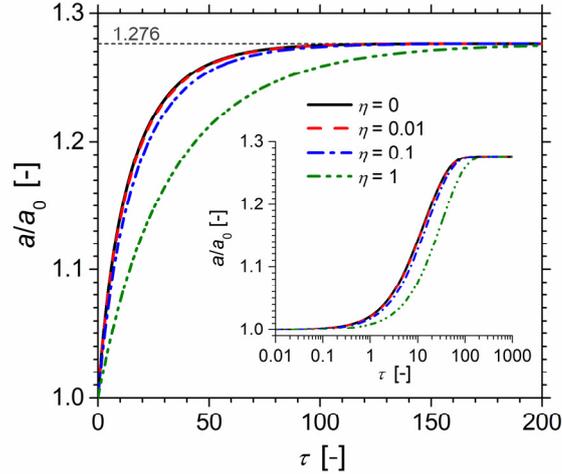

**Figure 2.** Influence of gas-liquid viscosity ratio ($\eta$) on the spreading of a spherical-cap droplet following the Cox theory for $\lambda = 10$.

### 4.2 Comparison of CLSVOF and phase field method

This test case is performed for $H = 2R_0$ and $\eta = 0.0005$ which corresponds to the dynamic viscosity of air ($\mu_G = 1.5 \cdot 10^{-5}$ Pa s). In the CLSVOF simulations, the grid is uniform with a mesh size $h = 0.1 \,\mu\text{m}$. Thus, $R_0$ is resolved by 500 mesh cells and the computational domain is discretized by 1000×1000 cells. For solution of the governing equations, the PISO algorithm is used in combination with the QUICK scheme. Time integration is performed implicitly by a first order upwind scheme with a fixed time step of $\Delta t = 1\,\mu s$.

The simulations with phaseFieldFoam are performed on a uniform grid as well. $R_0$ is resolved by 150 mesh cells. This corresponds to $300\times 300$ cells for the entire computational domain and a mesh size $h = 3.\overline{3}\,\mu m$. The Cahn number is set to $Cn = 0.0126$. The corresponding value of the interface thickness parameter is $\varepsilon = 5\,\mu m$. The range $-0.9 \leq C \leq 0.9$ is then resolved by about 6 mesh cells. The Peclet number is $Pe_C = 2494$ which corresponds to the mobility $\kappa = 2.5 \cdot 10^{-11}\,m^3 s/kg$. The time step $\Delta t = 0.2\,\mu s$ is constant during the simulation.

**Figure 3** shows the time evolution of the spreading radius (normalized by the initial value $a_0$) computed by the CLSVOF and PF methods. For both methods, the terminal spreading radius agrees well with the analytical value

$$\frac{a_e}{a_0} = \left(\frac{2 - 3\cos\theta_0 + \cos^3\theta_0}{2 - 3\cos\theta_e + \cos^3\theta_e}\right)^{1/3} \sin\theta_e = \sqrt[3]{\frac{2}{5}}\sqrt{3} = 1.276 \tag{30}$$

However, the spreading dynamics differs notably. Also shown in **Figure 3** are spreading curves of the Cox law. The latter involves the phenomenological parameter $\lambda = \ln(L/L_s)$ which is expected to be about $10$. To account for the uncertainty of this parameter, the spreading curves for $\lambda = 9$ and $11$ are displayed in **Figure 3** as well. For the CLSVOF method, the time-dependent spreading curve is in very good agreement with the Cox law for $\lambda = 11$ while that obtained with PFF is in agreement with $\lambda \approx 8$. For comparison, the simulation with PFF was also performed with gravity force ($g = 9.81\,\text{m/s}^2$). In this case, it is $a_e/a_0 = 1.281$ which is about 0.4% larger as compared to the case without gravity.

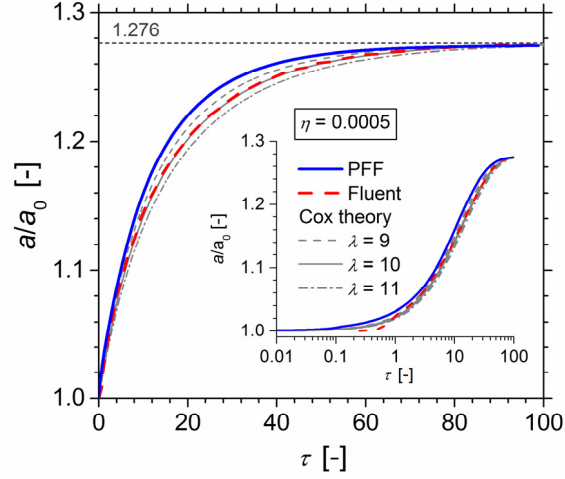

**Figure 3.** Time evolution of normalized spreading radius $a(\tau)/a_0$. Comparison of simulations results with Cox theory for different values of $\lambda$.

From the curves $a(\tau)/a_0$ displayed in **Figure 3**, the time evolutions of the capillary number

$$Ca(\tau) = \frac{\mu_\text{L} U}{\sigma} = \frac{a_0}{R_V} \frac{\text{d}}{\text{d}\tau}\left(\frac{a(\tau)}{a_0}\right) \tag{31}$$

and the Reynolds number

$$Re(\tau) = \frac{\rho_\text{L} R_V U}{\mu_\text{L}} = \frac{Ca(\tau)}{Oh^2} \tag{32}$$

can be computed. The maximum values occur for $\tau = 0$. While the maximum capillary number $Ca(\tau=0) \approx 0.48$ is less than unity, the maximum Reynolds number $Re(\tau=0) \approx 5.7$ is larger than unity. Thus, the requirement $Re < 1$ for the Cox-Voinov theory to be valid is not met in the initial stage of the computations.

The theory in Section 2 relies on the assumption that at each instant in time of the spreading process the drop forms a spherical cap. In **Figure 4**, the drop shapes computed by PFF for different instants in time are displayed. From the region close to the contact line it is evident that for $t = 0.5\,\text{ms}$ the drop shape differs from a spherical cap. A similar behavior is found in the CLSVOF simulations with Fluent. Due to the boundary conditions for the contact angle, the contact angle in the simulations immediately adapts from the initial value $\theta_0$ to the equilibrium value $\theta_e$ while away from the wall the interface remains at its initial state.

The described distortion of the interface can induce a capillary wave that travels up the drop and may even result in drop ejection.[27] In this context, it is useful to perform a simple estimation of the critical time $t_{cr}$ when the capillary wave is diminished so that one may expect agreement between the CFD results and Cox theory. The speed of a capillary wave with wave length $\Lambda$ is $U_{cw} \sim \sqrt{2\pi\sigma / \Lambda \rho_L}$. The time required for this wave to travel a distance $L_{cw}$ is

$$t = \frac{L_{cw}}{U_{cw}} \sim \sqrt{\frac{\rho_L \Lambda L_{cw}^2}{2\pi\sigma}} \tag{33}$$

For evaluation of $t_{cr}$ we assume that the capillary wave travels from the contact line to the drop apex and back so that $L_{cw} = \pi R_0$. As a rough approximation for the mean wave length of the capillary wave during this period we take one half of the distance from the contact line to the drop apex so that $\Lambda = \pi R_0 / 4$. Then Eq. (33) yields $t_{cr} \sim \sqrt{\pi^2 \rho_L R_0^3 / (8\sigma)} = 2.2\,\text{ms}$ which corresponds to $\tau_{cr} \sim 5.4$.

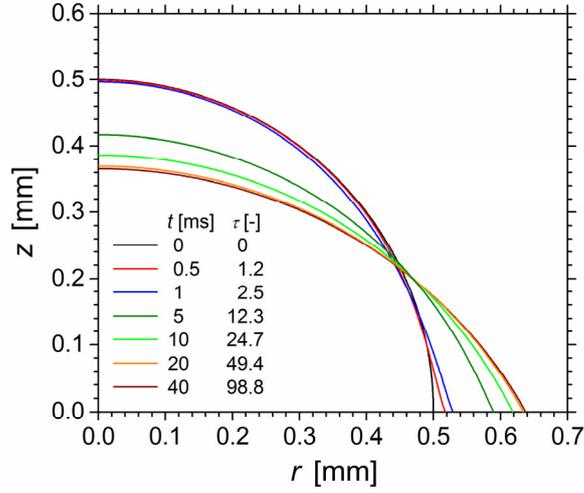

**Figure 4.** Droplet shapes for different instants in time (results of phaseFieldFoam).

Another possible reason why a good agreement between the simulation results and Cox theory may not be expected for short times is inertia, as the simulations start from a static state where the droplet is at rest. In the present study, the Ohnesorge number $Oh := \sqrt{\mu_L^2 / (\rho_L \sigma R_V)} = 0.291$ is somewhat smaller than unity, which indicates that inertia may have an influence in the initial stage of spreading. The theory for spherical-cap droplets in Section 2 does not consider the effect of inertia. For small values of $t$, the numerical spreading curves may thus be delayed as compared to those computed from Cox theory.

### 4.3 Effect of inertia

To investigate and quantify the effect of inertia, two additional phase-field simulations are performed for the same value of $Oh = 0.291$; once with and once without inertia. The computations are carried out with the AMPHI code that employs Galerkin finite elements on an adaptive triangular grid that adequately resolves the interfacial region. In this test case, $\varepsilon = 0.5\,\mu\text{m}$ and $Cn = 0.00126$ are used. Both parameters are thus ten times smaller than in phaseFieldFoam. The mobility $\kappa = 8.33 \cdot 10^{-11}\,\text{m}^3\text{s/kg}$ is about 3.3 times larger than in PFF. These changes result in a Peclet number $Pe_C = 74.8$ which is about 33 times smaller than in PFF. In order to minimize the influence of the domain size $H = 3R_0$ is used. The gas-liquid viscosity ratio is $\eta = 0.01$. According to **Figure 2**, the results for $\eta = 0.01$ should not differ from those for $\eta = 0.0005$. With these parameters and the adaptive finite element mesh, the interface is much thinner but much better resolved as compared to PFF. Consequently, the simulations with the AMPHI code fulfill the criteria for the sharp-interface limit of the phase field method[16], which is not the case for the PFF simulations.

The simulation results are shown in **Figure 5**. The influence of inertia is rather small but still notable, which is consistent with the value of $Oh$. As expected, inertia tends to slow down the spreading process in the initial stage but speeds it up in the final stage. Biance et al.[28] estimated the duration of the inertial regime as

$$t_{\text{inertial}} \sim \left(\frac{\rho_L \sigma R_V}{\mu_L^2}\right)^{1/8} \sqrt{\frac{\rho_L R_V^3}{\sigma}} \tag{34}$$

Though this relation was developed for completely wetting liquids, it may be useful for the present case of partially wetting liquids as well. When normalizing Eq. (34) by the capillary-viscous time scale, it follows

$$\tau_{\text{inertial}} := \frac{t_{\text{inertial}}}{t_{\text{ref}}} = K \cdot Oh^{-5/4} \tag{35}$$

where $K$ is an (unknown) non-dimensional pre-factor. In the present case it is $Oh^{-5/4} = 4.676$. To determine the proportionality factor $K$, we utilize the spreading curves with and without inertia displayed in **Figure 5**. Both curves intersect at $\tau = 5.861$ as indicated by the vertical line. Assuming that this intersection-time is representative for $\tau_{inertial}$ allows to determine the constant pre-factor in Eq. (35). Here, it follows $K = 1.253$. This value is of order unity, which supports the validity of the present approach.

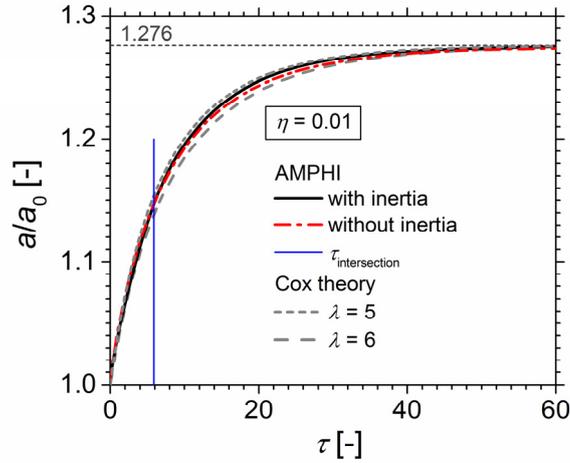

**Figure 5.** Time evolution of normalized spreading radius $a(\tau)/a_0$. Comparison of simulations results with and without inertia force (code AMPHI) with Cox theory.

The value of the normalized spreading radius at the cross-over time is $a(\tau_{inertial})/a_0 = 1.1465$. The corresponding instantaneous contact angle $\theta_{inertial}$ can be determined by solving Eq. (6) iteratively. Here one obtains $\theta_{inertial} = 73.51°$. Running the MATLAB script for Cox theory with this initial contact angle for different values of $\lambda$ yields a family of spreading curves beginning at $\tau_{inertial}$. **Figure 6** shows the corresponding spreading curves for $\lambda = 5$ and $\lambda = 6$ (with the vertical line denoting $\tau_{inertial} = 5.861$). The curve for $\lambda = 5$ almost overlaps with the computed spreading curve (case with inertia). This allows estimating the effective slip length for this case as $L_S \approx L \cdot \exp(-5) \approx 0.0067 R_V \approx 2.67 \mu m$.

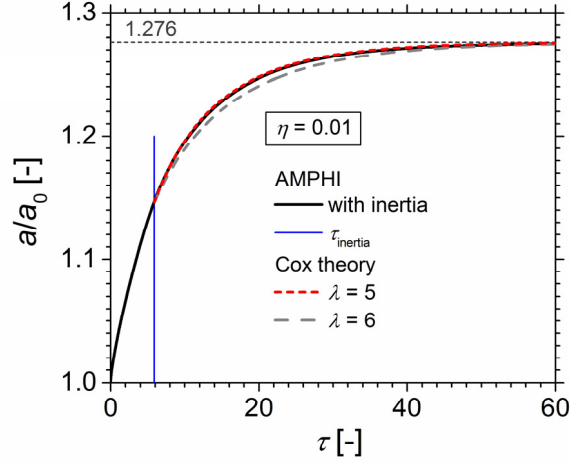

**Figure 6.** Time evolution of normalized spreading radius $a(\tau)/a_0$. Comparison of simulation result with inertia (code AMPHI) with Cox theory computed with $\theta_0 = \theta_{inertial} = 73.51°$.

## 4.4 Relation between slip length and diffusion length in phase field method

In phase field methods there exists in addition to $\varepsilon$ (which represents the length scale over which $C$ varies) a much larger length scale $L_D$ over which the Cahn-Hilliard diffusion takes place (and over which the chemical potential and the velocity vary). For non-matched viscosity ratios, Yue et al.[16] proposed the relation $L_D = \sqrt{\kappa \mu_{eff}}$ where $\mu_{eff} = \sqrt{\mu_L \mu_G} = \mu_L \sqrt{\eta}$ is an effective viscosity. They also introduced the non-dimensional group $S := L_D / L$. This length scale ratio is related to other non-dimensional groups as

$$S = \left( \frac{2\sqrt{2\eta}}{3} \frac{Cn \cdot Ca_{\text{ref}}}{Pe_C} \right)^{1/2} \tag{36}$$

where $Ca_{\text{ref}} := \mu_L U_{\text{ref}} / \sigma$ is a reference capillary number which is here equal to unity (due to the specific definition of $U_{\text{ref}}$ as capillary-viscous velocity scale). Yue et al.[16] argued that $L_D$ corresponds to the slip length $L_S$ in Cox theory so that $\lambda = \ln(L/L_S) \approx \lambda_D := \ln(L/L_D) = \ln(S^{-1})$. Combining this relation with Eq. (36) yields

$$\lambda_D = \frac{1}{2} \ln\left( \frac{3}{2\sqrt{2\eta}} \frac{Pe_C}{Cn} \right) \tag{37}$$

For a given viscosity ratio, Eq. (37) can be used to estimate the effective slip parameter $\lambda_D$ from the numerical parameters $Cn$ and $Pe_C$. For the simulation with phaseFieldFoam one obtains $\lambda_D = 7.98$ which is in surprisingly good agreement with **Figure 3**. For the simulation with AMPHI, it is $\lambda_D = 6.67$. This value is clearly larger than the value $\lambda \approx 5$ suggested by **Figure 6**. No measures have been undertaken so far to get a better agreement between $\lambda_D$ from Eq. (37) and $\lambda$. However, taking $\lambda_S = 2.5\lambda_D$ as suggested by Yue et al.[16] instead of $\lambda_S = \lambda_D$ already yields $\lambda_{D*} = 5.76$ for the AMPHI results which is in better agreement with $\lambda \approx 5$. Furthermore, the function relationship $\mu_{\text{eff}} = \sqrt{\mu_L \mu_G} = \mu_L \sqrt{\eta}$ has been proposed in an adhoc manner without deeper scientific foundation.

The semi-analytical procedure developed in this paper is of special significance for numerical methods for moving contact lines. For diffusive interface methods, it offers the opportunity to refine Eq. (37) and establish a clear quantitative relationship between the mobility (or Peclet number) and the effective slip length $L_S$. This relation should be unique and independent of discretization and code implementation; at least as long as the diffuse interface is well resolved (which is not the case for the PFF simulations). The only uncertainty is the effective viscosity. Establishing such a relation for the phase-field method requires considerable computations for a wide parameter space. This is beyond the scope of this paper but will be considered in future work.

Finally it is remarked, that for sharp interface methods such as the VOF and CLSVOF methods there is no such relation as Eq. (37). However, Renardy et al.[29] related the effective slip length in their VOF method to the grid resolution near the wall. The procedure developed in the present paper may be used to establish quantitative relationships between effective slip length and grid resolution for sharp interface methods as well. For convenience, the procedure is shortly summarized in the following subsection.

### 4.5 Procedure for estimating the effective slip length from spreading curves

In this paper, a semi-analytical procedure is presented that allows estimating the effective slip length $L_S$ of sufficiently small droplets from radius-over-time curves obtained either from experiments or from numerical computations. It consists of the following five steps:

1. Check if the Eötvös number is sufficiently small so that gravitational effects are negligible and the spherical-cap assumption is justified. Results of numerical studies from literature[12, 30, 31] suggest that this is the case for $Eo \leq 0.1$.

2. Compute the Ohnesorge number and estimate (for $Oh = \sqrt{\mu_L^2/(\rho_L \sigma R_V)} < 1$) from Eq. (35) the duration of the inertial regime $\tau_{\text{inertial}} = K \cdot Oh^{-5/4}$ by using either the present value of the pre-factor ($K = 1.253$) or simply $K = 1$.

3. Determine from the instantaneous (experimental or numerical) value of the spreading radius at the inertial time $a(\tau_{\text{inertial}})$ the corresponding instantaneous contact angle $\theta_{\text{inertial}}$ by solving Eq. (6) iteratively.

4. Run the present MATLAB script (see supplemental material) with $\theta_0 = \theta_{\text{inertial}}$ to determine the spreading curves of the general Cox law for different values of $\lambda = \ln(L/L_S)$.

5. Determine the value $\lambda = \lambda_{\text{fit}}$ which best fits to the experimental or numerical spreading curve for $\tau \geq \tau_{\text{inertial}}$ and estimate the effective slip length by relation $L_S \approx L \cdot \exp(-\lambda_{\text{fit}})$ with an appropriate value of the macroscopic length scale ($L$), e.g. $L = R_V$ as used here.

# 5   CONCLUSIONS

Experimental or computational results on spreading processes naturally emerge as base-radius over time curves. In this paper, a semi-analytical method is presented that allows directly comparing such curves with those arising from the general Cox relation between apparent contact angle and contact line speed. The advantage of the method is that neither a measurement of the apparent contact angle nor a computation of the contact line speed by differentiation of the spreading radius is necessary. The derived mathematical relationship is only valid if gravitational effects are negligible so that the drop forms at each instant in time a spherical cap. The spreading dynamics according to the Cox law is found to become independent from the gas viscosity when the gas-liquid viscosity ratio is below about $0.01$.

The presented procedure is useful for a straightforward comparison of experimental or computational results on droplet spreading in the capillary-viscous regime with Cox theory. To exclude the effect of inertia, which may invalidate the comparison with Cox theory for the initial stage of spreading, a non-dimensional time scale ($\tau_{\text{inertial}}$) is derived when the effect of inertia has diminished. This time scale is related to the Ohnesorge number by a power law. For times larger than $\tau_{\text{inertial}}$, comparison of experimental or computational spreading curves with Cox theory allows fitting of $\lambda = \ln(L/L_S)$ and thus determining the effective slip length.

The semi-analytical procedure developed in this paper is of potential benefit for the advancement of numerical methods for moving contact lines. For phase-field methods, it offers the opportunity to establish a quantitative relationship between the effective slip length and the mobility (which is usually treated as a numerical parameter rather than a physical one). Establishing such a relation requires considerable computations for a wide parameter space, a task that will be considered in future work.

For spreading processes where gravitational forces are not negligible, the drop shape will deviate from a spherical cap. The present method may be extended to such situations by introducing an appropriate (yet unknown) functional relationship for the instantaneous drop shape (e.g. in terms of the Eötvös number) into Eq. (6).

## ACKNOWLEDGEMENTS

M.W. and X.C. acknowledge the financial support by Helmholtz Energy Alliance "Energy-efficient chemical multiphase processes" (HA-E-0004). X.C. is grateful for the Research Travel Grant funded by the Karlsruhe House of Young Scientists (KHYS) to support his research stay at Virginia Tech. P.Y. acknowledges the support by NSF-DMS (Grant No. 1522604). The simulations with phaseFieldFoam are obtained using computational resources of bwHPC

(http://www.bwhpc-c5.de/index.php), funded by the Ministry for Education and Research (BMBF) and the Ministry for Science, Research and Arts Baden-Wuerttemberg (MWK-BW). M.W. and X.C. are especially grateful to H. Marschall for his significant support in the development and implementation of the phase-field method in OpenFOAM®. The authors also acknowledge useful discussions with M. Ben Said (M.W., X.C., H.A.) and D. Bothe (M.W.).## REFERENCES